%% file: split_society.tex
\documentclass[
  aps,
  pre,
  twocolumn,
  amsmath,
  amssymb,
  groupedaddress
  ]{revtex4-2}

\usepackage[utf8]{inputenc}
\usepackage[T1]{fontenc}
\usepackage{graphicx}
\usepackage{xcolor}
\usepackage{hyperref}
\definecolor{hyperrefcolor}{HTML}{1F77B4}
\hypersetup{allcolors=hyperrefcolor,colorlinks=true}

\newcommand{\numberOfVoters}{94}
\newcommand{\fieldDataDate}{{1990}}
\newcommand{\eqmgn}{s^\ast}
\newcommand{\figlet}[1]{\lowercase{\textsf{#1}}}

\begin{document}

\title{Dislike of general opinion makes for tight elections}

\author{O.~Devauchelle}
\affiliation{Université Paris Cité, Institut de Physique du Globe de Paris, 1 rue Jussieu, CNRS, F-75005 Paris, France}
\email{devauchelle@ipgp.fr}

\author{P.~Szymczak} 
\affiliation{Institute of Theoretical Physics, Faculty of Physics, University of Warsaw, Poland}

\author{P.~Nowakowski}
\affiliation{Group for Computational Life Sciences, Division of Physical Chemistry, Ruđer Bošković Institute, Zagreb, Croatia and Max Planck Institute for Intelligent Systems, Stuttgart, Germany}

\begin{abstract}
In modern democracies, the outcome of elections and referendums is often remarkably tight. The repetition of these divisive events are the hallmark of a split society; to the physicist, however, it is an astonishing feat for such large collections of diverse individuals. Many sociophysics models reproduce the emergence of collective human behavior with interacting agents, which respond to their environment according to simple rules, modulated by random fluctuations. A paragon of this class is the Ising model which, when interactions are strong, predicts that order can emerge from a chaotic initial state. In contrast with many elections, however, this model favors a strong majority. Here, we introduce a new element to this classical theory, which accounts for the influence of opinion polls on the electorate. This brings about a new phase in which two groups divide the opinion equally. These political camps are spatially segregated, and the sharp boundary that separates them makes the system size-dependent, even in the limit of a large electorate. Election data show that, over the last 30 years, countries with more than about a million voters often found themselves in this state, whereas elections in smaller countries yielded more consensual results. We suggest that this transition hinges on the electorate's awareness of the general opinion.
\end{abstract}

\date{\today}

\maketitle


\begin{figure}[t]
  \centering
  \includegraphics[width=.99\linewidth]{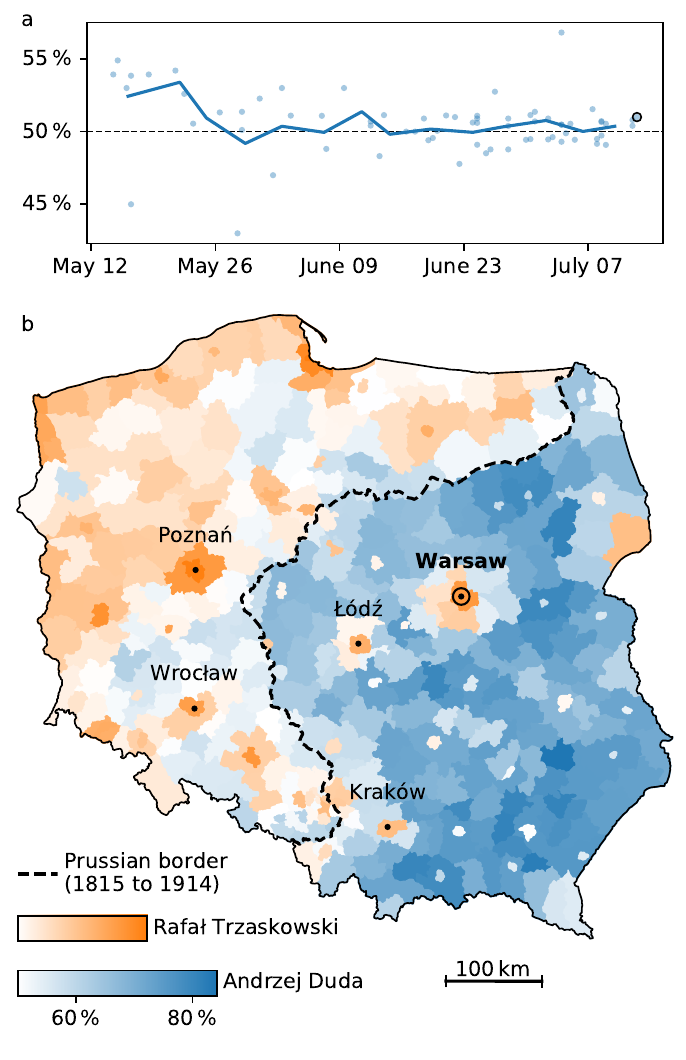}
  \caption{Polish presidential election in 2020.
  \figlet{A} Opinion polls before the election. Source: \citet{enwiki:1038382064}. Blue dots: opinion in favor of Andrzej Duda (App.~\ref{app:op_polls}). Solid blue line : average over 4-days bins (App.~\ref{app:fluctuations}). Blue circle: election result.\label{fig:Polish}
  \figlet{B} Map of the election result in Polish counties (\emph{powiat}). Source: National Electoral Commission of Poland \cite{polishelectoralcom}. The black dashed line shows the Prussian border in 1890 \citep{kashin}, virtually unchanged from 1815 to 1914 \citep{zarycki2015electoral}.\label{fig:Polish_map}
  }
\end{figure}

In May 2016, after a divisive campaign, the United Kingdom voted to leave the European Union, with a majority of 51.9$\,$\%---a close call indeed \citep[e.g.][]{arnorsson2018causes}. Does this mean that, as chance would have it, about half of British voters were in favor of leaving, while the other half opposed? Opinion polls before the vote suggest a different picture. In January 2011, Remainers where ahead of Brexiteers by about 20 percentage points (pp), but this gap fluctuated widely in the following polls, while slowly shrinking on average \citep{enwiki:1038382064}. At some point in 2013, the fluctuations became comparable to the gap itself, and the two curves that would decide the fate of the UK started to routinely cross---the electorate had reached a polarized opinion.

In fact, given a binary choice, the opinion of modern democracies often splits itself into remarkably even parts \citep{levin2021dynamics}. During the 2020 presidential election in Poland, for instance, opinion surveys showed a similar evolution: over the last two weeks before election day, the polls relaxed towards equality (Fig.~\ref{fig:Polish}\figlet{A}). The actual election yielded only a small margin to the winner, Andrzej Duda ($51.0\,$\% of the expressed votes).
What leads a population to distribute its votes so evenly between two political options? The null hypothesis, of course, is that every citizen flips a coin to pick their favorite candidate.

Fig.~\ref{fig:Polish_map}\figlet{B} shows the geographical distribution of votes during the same election. Essentially, the votes distribute themselves into two main clusters: the Eastern part of the country voted for Andrzej Duda, whereas the West voted for Rafał Trzaskowski, with smaller clusters around cities. We take this clustering,  a common feature of electoral maps \citep[e.g.][]{araujo2010tactical,sobkowicz2016quantitative,arnorsson2018causes}, as a rebuttal of the null hypothesis.

An alternative route is to represent an electorate as a collection of interacting agents. There is little doubt, indeed, that our decisions depend on the opinion of others, and on how closely we relate to them \citep{latane1981psychology}. Right before the era of desktop computers, for instance, \citet{schelling1969models} reproduced the formation of segregated neighborhoods with deterministic agents who choose their next abode based on simple rules. The agents distribute themselves on a grid, until an equilibrium is reached, which features clusters of homogeneous population. These clusters, however, remain small, because the agents' decisions are strictly deterministic, and based on local information \citep{stauffer2007opinion}.

In the celebrated voter model \cite{cox1986diffusive}, the opinion $s$ of each citizen can take two values ($+1$ or $-1$) and, at each time step, every voter transmits its opinion to a randomly chosen neighbor of theirs. At first, it certainly seems rough (and slightly degrading) to reduce a citizen's political views to a two-state variable submitted to such simplistic evolution rules. The collective behavior of these mechanical citizens, however, is not trivial. In fact, it depends crucially on the topology of the network that connects the voters. When only neighbors are connected, rough clusters form and diffuse through the network in dimension two or less, but not in higher dimensions \cite{dornic2001critical,miguel2005binary,castellano2009nonlinear}.  When they exist, these clusters coarsen, until one invades the entire domain, and the population eventually reaches unanimity. This is also true for the related majority-rule model, wherein voters adopt the dominant opinion of a random set of their neighbors \cite{PhysRevLett.90.238701}.

This relaxation to consensus is not surprising if one assumes that voters tend to align their opinion with their neighbors (or friends). This is exactly what happens in the Ising model, which was initially designed to represent a magnet, and later became the archetypal model of phase transition \citep{Ising1925,brush1967history}. When thermal fluctuations are small, the spins of a magnet align with each other, and thus induce a macroscopic magnetic field. This ferromagnetic behavior disappears above the Curie temperature, as disorder takes over the system. The analogy with the social behavior of humans (and fish, for that matter) was recognized early \citep{callen1974theory}, but the original Ising model, like the voter model, cannot evenly split an electorate (unless the null hypothesis prevails).

One way to circumvent this limitation is to make voters influence their neighbors' in a deterministic way \citep{sznajd2000opinion,bernardes2002election}. A population of such voters ultimately reaches an ordered equilibrium: either a complete consensus, or a gridlock wherein every voter opposes their neighbors. In the last scenario, the opinion is evenly split, but the initial state needs to be finely tuned to ensure this outcome.

To prevent a single ordered phase to invade the entire population, \citet{nowak1990private} randomly alters the bond between two voters after their opinion has flipped. This disconnects some groups of voters from the majority, and allows them to hold their views. The propagation of politically-oriented information through a community can also break cross-ideology ties; this rewiring is visible in the network data of social media \citep{tokita2021polarized}. Another way to maintain some cultural diversity is to strengthen the influence of like-minded individuals over each other, a processed called ``homophily'' \citep{axelrod1997dissemination,miguel2005binary,holme2006nonequilibrium,baumann2020modeling,korbel2023homophily,liu2023emergence}. A social network can even evolve on its own, regardless of its users' opinion, as recommendation algorithms steers them toward isolated communities \citep{santos2021link}. These mechanisms, however, do not favor any special partition of the opinion. For this to happen, the voters need to be either marshaled, or informed about the general opinion.

Opinion polls, when publicly available during an election, can affect its outcome, for instance by enticing some electors to cast their vote in a head-to-head election, when they might not have cared to otherwise \citep{restrepo2009modeling}. Tactical considerations informed by opinion polls also enter electoral dynamics \citep{araujo2010tactical}. When there are more than two candidates, one might indeed prefer to use their vote to evict a candidate they strongly dislike, and thus vote for a contender that appears more likely to win than their favorite. This might explain why two candidates often take over the election, while the others are marginalized.

In binary elections, there can be no tactical voting; opinion polls can still influence a voter's decision, but not for strategic reasons. Clearly, if all voters were purely conformist, the polls would drive them towards unanimity. Nonconformists, however, can drastically change the propagation of trends and opinions in a population \cite{dodds2013limited,touboul2019hipster,juul2019hipsters}. Here, we add an element of nonconformity to the Ising model, by assuming that voters tend to \emph{oppose} the general opinion, while remaining faithful to their friends. In the words of social psychology, they have a negative attitude toward the winning camp \citep{dalege2017network}.

\citet{pham2022empirical} and \citet{korbel2023homophily} recently introduced this tension between homophily and heterophily (members of a group tend to oppose those of other groups) in their glass-spin models. Evolving the connectivity of the population according to the affinity between two individuals, \citet{korbel2023homophily} found that the population spontaneously fragments into groups, the size of which follows a realistic distribution.

In the model we propose here, heterophily also plays a key role, but every voter experiences it towards the entire population. The average opinion, communicated by the media, is perceived as the opinion of others---which every voter would like to oppose. To fix ideas, we could call this negative attitude towards the average opinion an anti-establishment feeling, or an ingrained guard against the rule of the majority. However, we shall not speculate further about its psychological or cultural origin; rather, we shall investigate its mathematical consequences, and compare them to actual elections.

\section{Herding and dislike of general opinion}

Inspired by \citet{nowak1990private} and \citet{araujo2010tactical}, we represent an electorate with a population of $N$ voters, each holding one of two opposite opinions:
\begin{equation}
  s_i = \pm 1, \quad i \in \{0,1,\dots,N-1\} \, .
\end{equation}
The state of a voter is thus binary, like the spins of the Ising model. With this definition, the general opinion is the average of individual opinions:
\begin{equation}
\bar{ s } = \dfrac{1}{N} \sum_{i=0}^{N-1} s_i \, ,
\end{equation}
and the fraction of the electorate that supports opinion $+1$ is $(\bar{s}+1)/2$.

We now define rules to evolve the electorate. To express them, we introduce a Hamiltonian ${\cal H}(\bf{s})$, with the underlying idea that states for which ${\cal H}$ is large are unlikely to occur spontaneously ($\bf{s}$ is the vector of dimension $N$ that represents the state of the electorate). We also want ${\cal H}$ to account for the influence of a voter's neighbors on their  opinion, and for their sensitivity to publicly-available poll results. A simple expression for the Hamiltonian is
\begin{equation}
  {\cal H} = 
  - \sum_{i,j=0}^{N-1} J_{ij} s_i s_j + {N\varepsilon \bar{s}^2} \, ,
  \label{eq:H}
\end{equation}
where $J_{ij}$ corresponds to the influence of voter $j$ on voter $i$, and $\varepsilon$ is the sensitivity of a voter to general opinion.

\begin{figure}[t]
  \centering
  \includegraphics[width=.99\linewidth]{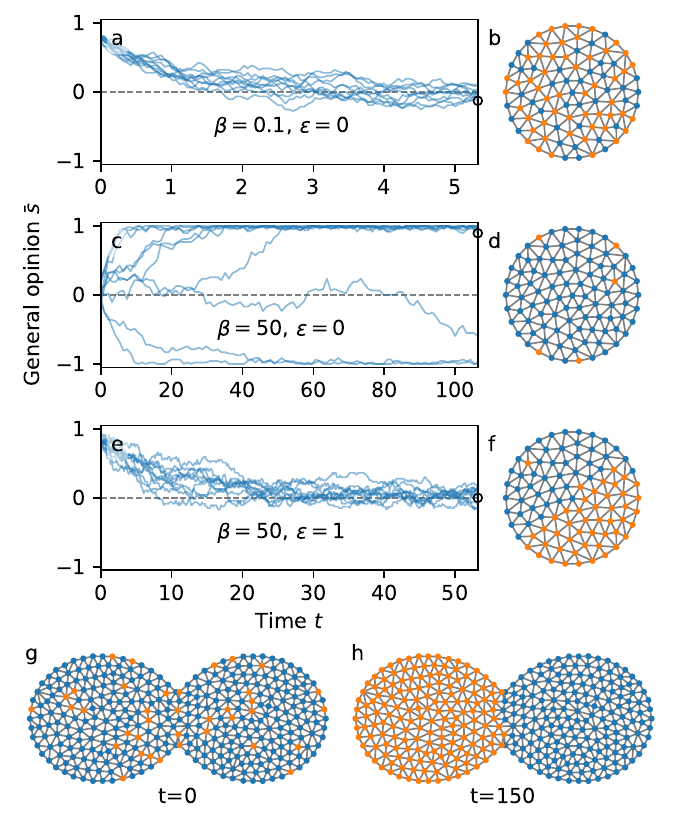}
  \caption{
    \figlet{A}--\figlet{F} Numerical simulations on a {\numberOfVoters}-voters triangular mesh.
    \figlet{A}, \figlet{C}, \figlet{E}: Evolution of the average opinion.
    \figlet{A}, \figlet{C}, \figlet{E}: Examples of final state. Corresponding point shown on \figlet{A}, \figlet{C} and \figlet{E} with black circles.
    \figlet{G}, \figlet{H}: Initial and final opinion of 335 voters distributed on a narrow-necked mesh ($\beta=30$, $\varepsilon=1$). \figlet{A}, \figlet{E}, \figlet{G}: Initial opinions are randomly distributed with 90$\,$\% of blue votes; \figlet{C}: 50$\,$\% of blue votes. \label{fig:peanut_bag} \label{fig:evolution}}
\end{figure}

The first term in Eq.~\eqref{eq:H} is common to all Ising models; here, it represents social impact \citep{nowak1990private,araujo2010tactical}. We assume that $J_{ij}=J_{ji}=1/2$ when voters $i$ and $j$ are socially connected, and $J_{ij}=0$ otherwise. The structure of the electorate can thus be represented by a lattice, in which the nodes represent voters, and edges represent social connections (Fig.~\ref{fig:evolution}\figlet{B}, for instance). Since $\mathbf{J}$ is a nonnegative matrix, the corresponding term in the Hamiltonian favors unanimity: the energy of the population (i.e. the value of the Hamiltonian) decreases when connected voters agree.

The second term in Eq.~\eqref{eq:H}, which reads
\begin{equation}
\varepsilon N  \bar{s}^2 =
\varepsilon \bar{s} \sum_{i=0}^{N-1} s_i =
\dfrac{\varepsilon}{N} \sum_{i,j=0}^{N-1} s_i s_j
\, ,
\end{equation}
is unusual, although similar ones have appeared in some spin-glass models \citep[notably in the
cost function of computer-chip layout,][]{kirkpatrick1983optimization}. Here, it represents the impact of opinion polls on the electorate. We assume $\varepsilon$ to be positive, which means that, although voter $i$ tends to agree with their friends and neighbors, they prefer to oppose general opinion, in proportion to $-\varepsilon \bar{s}$. Equivalently, one can say that every possible pair of voters is coupled with a small negative coefficient $-\varepsilon/N$.

As a simple limit case, we may consider an entirely disconnected population (${\bf J}$ vanishes). The Hamiltonian then depends on the state of the electorate only through $\bar{s}^2$. Such a population will tend to distribute its votes evenly to minimize its energy ($\bar{s}$ will be close to zero). To the contrary, a well-connected population with no sensitivity to opinion polls ($\varepsilon=0$) will tend to vote uniformly when fluctuations are weak, although disconnected groups might form clusters of opposed opinion. In both cases, however, random fluctuations can perturb the system away from the low-energy states.

We now return to a non-trivial Hamiltonian, which represents voters that are sensitive both to their neighbors' opinion and to general opinion. To represent the influence of neighboring voters on each other, we need to specify the connectivity matrix ${\bf J}$. In line with earlier models \citep{cox1986diffusive,nowak1990private}, we assume that voters are distributed on a two-dimensional lattice, such as the one of Fig.~\ref{fig:evolution}\figlet{B}---a crude representation of electoral geography. As a result, the connectivity matrix is sparse: voters that are far away from each other on the lattice do not interact directly. The only long-range interactions in the present model are those that are mediated by opinion polls, through which each voter is connected to the electorate as a whole. A more realistic connectivity, perhaps, would involve a scale-free distribution of interactions \citep{barabasi1999emergence,bernardes2002election,grabowski2006ising} but, to keep things intuitive, we limit the present investigation to networks in which connectivity essentially maps onto geometrical distance.

Let us assume that each new issue of an opinion poll triggers a debate in the population of voters. As a result of this debate, some voter $i$ changes their mind under the combined influence of their neighbors and of the new poll. How likely is the transition $s_i \mapsto - s_i$? Obviously, this likelihood should decrease with the associated energy gain,
\begin{equation}
  \Delta E_i = {\cal H}({\bf s}) -{\cal H}( {\bf T}_i \cdot {\bf s} ) \, ,
\end{equation}
where the linear operator ${\bf T}_i$ realizes the transition $s_i \mapsto - s_i$. In our model, $\Delta E_i$ is the only quantity that systematically influences a voter.

Of course, we cannot model a voter's decision in a fully deterministic way. Instead, we treat the complicated dynamics of electoral decision-making as random fluctuations that are biased towards negative $\Delta E_i$. We represent the strength of this bias with a positive parameter, $\beta$, and assume that the probability $p_i$ that voter $i$ change their mind is a decreasing function of $\beta \Delta E_i$. Specifically, we choose the algorithm of \citet{glauber1963time} to evolve the electorate: at each step, we randomly pick a voter $i$ and flip their vote, also randomly, with probability $p_i$. This probability decreases exponentially with $\beta \Delta E_i$:
\begin{equation} 
  p_i = \dfrac{\exp(-\beta \Delta E_i) }{1 + \exp(-\beta \Delta E_i)} \, .
  \label{eq:Glauber}
\end{equation}
As visible in this expression, the units of the Hamlitonian do not really matter, provided the product $\beta {\cal H}$ is dimensionless.

We then pick another voter, and repeat the procedure; this loop defines an algorithm for the noise-driven evolution of the electorate. A natural definition for time is then $t = n_G/N $, where $n_G$ is the number of iterations. By this definition, at time $t$, each voter has been picked $t$ times on average.

After many iterations ($t\to\infty$), Glauber dynamics drives any Hamiltonian system toward the Boltzmann distribution, wherein the probability of a given state $\bf s$ is proportional to $ \exp(-\beta {\cal H}(\bf{s}))$. The analogy with classical thermodynamics is now obvious, and justifies our calling ``energy'' the value of the Hamiltonian. Similarly, we call $1/\beta$ the ``temperature'', with the understanding that it represents the propensity of a voter to follow their own judgment rather than the opinion of others.

For illustration, when the temperature vanishes ($\beta\to\infty$), the electorate is deterministic: voter $i$ automatically changes their mind when $\Delta E_i$ is negative, and sticks to their opinion otherwise. Conversely, an infinite temperature ($\beta \to 0$) means that every voter chooses their opinion based only on their own judgment (at random, in the present model). In between these two extremes, an intermediate value of $\beta$ portends richer dynamics.

\section{Split society}

We are now ready to numerically simulate  the evolution of our model. To do so, we first create a two-dimensional, triangular lattice with {\numberOfVoters} voters (Fig.~\ref{fig:evolution}\figlet{B}, App.~\ref{app:numerical_simulations}). Each edge in this lattice corresponds to a positive element in the connectivity matrix $\bf{J}$---two socially connected voters.

We then run a first series of simulations without any opinion poll ($\varepsilon=0$, App.~\ref{app:numerical_simulations}). When the temperature is large enough ($1/\beta=10$, for instance), we find that the opinion relaxes to about 50$\,$\% (Fig.~\ref{fig:evolution}\figlet{A}). This is not surprising, since voters are then indifferent to both their neighbors and general opinion---at large temperature, the model boils down to the null hypothesis. As it turns out, the general opinion then behaves much like actual opinion polls (Fig.~\ref{fig:Polish}\figlet{A}), but the opinions of individual voters are scattered randomly across the network (Fig.~\ref{fig:evolution}\figlet{B}). The absence of any structure in this distribution indicates that the connectivity of the electorate does not really matter at high temperature.

The picture changes radically when we decrease the temperature ($1/\beta = 0.02$, Fig.~\ref{fig:evolution}\figlet{C},\figlet{D}), that is, when voters become unlikely to change their mind if that means opposing their neighbors. As a result, even if we start the simulation with evenly distributed votes, the population now relaxes to unanimity. The resulting consensus can be either of the two options, but most voters then think alike. Such a population is ruled by the majority; only seldom does some isolated voter dare opposing the general opinion. If we want to represent a population in which social connections matter, but which does not give in to the majority altogether, we still lack an ingredient.

We now switch on the influence of opinion polls, by setting $\varepsilon = 1$, while maintaining a low temperature ($1/\beta = 0.02$, Fig.~\ref{fig:evolution}\figlet{E},\figlet{F}). Like in the high-temperature case, the average opinion relaxes to evenly-split votes, although at the slower pace dictated by a low temperature. Again, the evolution of the opinion resembles that of Fig.~\ref{fig:Polish}\figlet{A}. However, the distribution of individual opinions in this population is entirely different from that of Fig.~\ref{fig:evolution}\figlet{B}: Like-minded voters gather into two clusters of comparable size. The two camps are virtually unanimous, and their sizes match each other almost perfectly---the epitome of a split society. The rest of the paper focuses on the existence and the properties of this new, split-society phase.

\section{Vote pattern and social rift}

Although the coin-flipping electorate and the interacting, poll-sensitive one both relax to 50$\,$\% (figures~\ref{fig:evolution}\figlet{A}, \figlet{E}), the corresponding patterns on the connectivity lattice look entirely different (figures~\ref{fig:evolution}\figlet{B}, \figlet{F}). These patterns, of course, are only visible to an all-seeing observer, but an individual voter would nonetheless be able to tell them apart, based only on local perceptions.
One measure of these perceptions is the clustering coefficient, $c$, which is the average proportion of a voter's neighbors that share their opinion. At high temperature, only half of them agree ($c=0.5 \pm 0.01$ in the simulation, Fig.~\ref{fig:evolution}\figlet{B}). Conversely, in a split society, every voter is surrounded with like-minded neighbors, with barely any contact with political opponents ($c=0.9 \pm 0.02$)---except for those voters who find themselves on the boundary between the two camps (Fig.~\ref{fig:evolution}\figlet{F}).

Where does this boundary lies? On the disk lattice of Fig.~\ref{fig:evolution}\figlet{F}, it often settles along a diameter. To fix its position, we generate a peanut-shaped mesh, in which a narrow neck separates the network into two equal parts (Fig.~\ref{fig:peanut_bag}\figlet{G},\figlet{H}). This configuration creates a social rift: Voters are better connected to their own side of the network than to the opposite side. This rift exists prior to the election process; we now investigate how it expresses itself in the election result.

Starting the simulation with a biased opinion ($90\,$\% of blue votes initially, $\beta=30$ and $\varepsilon=1$), we find that clusters of like-minded voters appear and begin to merge (Fig.~\ref{fig:peanut_bag}\figlet{G},\figlet{H}). Over time, these clusters tend to split the population into two compact camps of equal size, and the boundary that separates them places itself right across the neck of the mesh. This location, of course, minimizes the length of the interface, and therefore minimizes the energy of the population; in an hourglass filled with two immiscible fluids, surface tension would favor the same configuration.

Clearly, the symmetric mesh of Fig.~\ref{fig:peanut_bag}\figlet{G},\figlet{H} favors the partition of the electorate into two equal camps. The electorate's connectivity and opinion polls thus concur to bring the average opinion to 50$\,$\%. An assymetric mesh, with one side larger than the other, would likely push the average opinion away from 50$\,$\%---an example of the connectivity's influencing an election (we shall not investigate this phenomenon here). In a real population, however, one can expect that the lines of weak connectivity are many, and therefore that the system can find a path along some of those lines that splits the electorate into two equal parts.


At the scale of a country, social connections pertain, at least partly, to geography.
We can thus expect that the spatial distribution of votes during an election bears the mark of social connections. In 1815, for instance, Poland was re-partitioned between Prussia, Austria and Russia, and this partition still lingers on in Polish society \cite{bukowski2019history}. If we superimpose the 1815 border to the 2020 electoral map (dashed line in Fig.~\ref{fig:Polish_map}\figlet{B}), we find that they match remarkably well. This phenomenon occurred in many Polish elections, and is often attributed to cultural and political differences between the three empires \citep{zarycki2015electoral,grabowski2019determinants}.

The present model suggests a distinct interpretation, whereby the social structure of present-day Poland bears the mark of historical borders, across which people are less likely to be connected---as visible, for instance, in marriage statistics \citep{grosfeld2015cultural,sleszynski,Sutowski2021}. Like in the narrow-necked lattice of Fig.~\ref{fig:peanut_bag}\figlet{H}, the old borders would then be a favorable location for the interface between political camps. The same mechanism might also account for the opposition between cities and rural areas around them (Fig.~\ref{fig:Polish_map}\figlet{B}). This interpretation is entirely speculative at this stage, but we take this resemblance with reality as encouragement.

\section{Fluctuation-induced phase transition}\label{sec:fluct_phase_trans}

\begin{figure}[t]
  \centering
  \includegraphics[width=.99\linewidth]{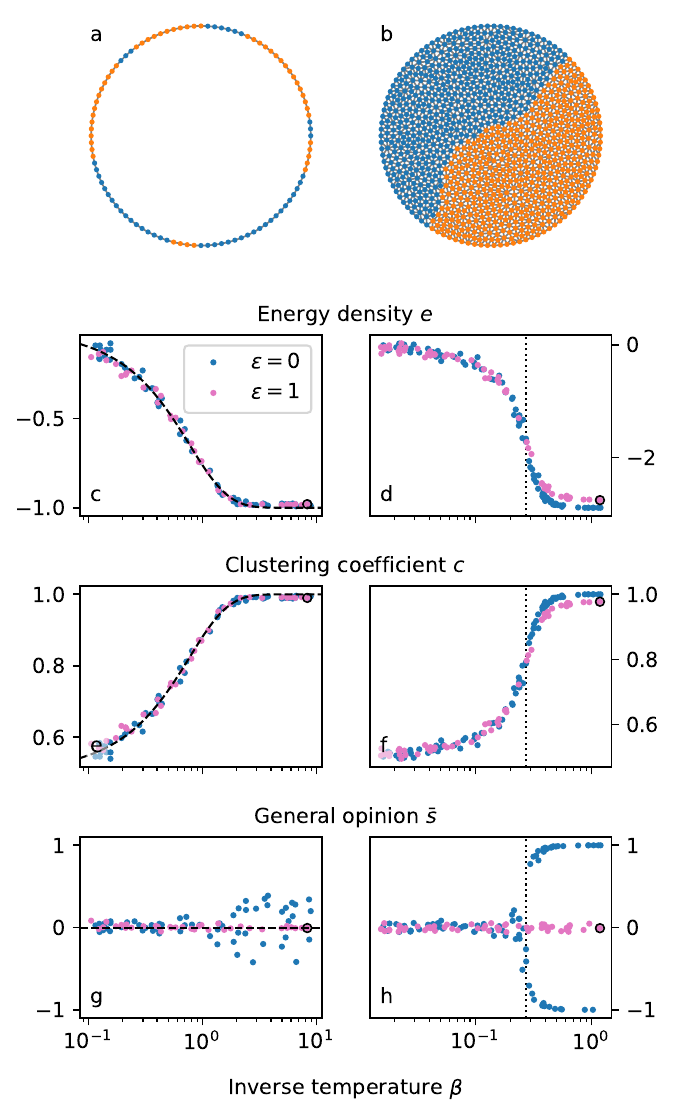}
  \caption{Fluctuation-induced transition. Left column: periodic one-dimensionnal lattice ($N=1000$, one out of ten nodes shown). Right column: triangular lattice ($N=1046$).
    \figlet{A}, \figlet{B} Examples of equilibrium states.
    \figlet{C}--\figlet{H} Final stage of the numerical simulation. Blue dots: no influence of the polls ($\varepsilon=0$), pink dots: polls matter ($\varepsilon=1$). Black circles correspond to states shown in \figlet{A}, \figlet{B}. Dashed black lines in \figlet{C}, \figlet{E} and \figlet{G} show the exact solution \citep[$\varepsilon=0$,][]{Ising1925}. The dotted black lines in  \figlet{D}, \figlet{F} and \figlet{H} show the critical temperature for a regular, infinite, triangular lattice, $\beta_c=\log(3)/4$.
  \label{fig:phase_transition}}
\end{figure}

The Ising model was introduced to explain the phase transition that temperature induces in ferromagnetic materials \citep{Ising1925,brush1967history}; we therefore expect that the present election model undergoes a similar transition.

We begin with the one-dimensional, periodic network shown in Fig.~\ref{fig:phase_transition}\figlet{A}; when $\varepsilon$ vanishes, this is the original Ising model in one dimension. It has a simple analytical solution which, in the thermodynamic limit ($N \to\infty$), shows no singular phase transition (black dashed lines in Fig.~\ref{fig:phase_transition}\figlet{C},\figlet{E},\figlet{G}). As they should, our simulations conform to this theory (blue dots in Fig.~\ref{fig:phase_transition}\figlet{C},\figlet{E},\figlet{G}): as $\beta$ increases, the energy density $e$ (total energy divided by $N$) decreases continuously, the clustering coefficient $c$ increases, and the general opinion $\bar{s}$ fluctuates about zero.

When the opinion-polls term is introduced ($\varepsilon=1$), not much changes, although the fluctuations of the average opinion at low temperature are reduced (pink dots in Fig.~\ref{fig:phase_transition}\figlet{C},\figlet{E},\figlet{G}). The simulations therefore suggest that, in one dimension, temperature induces no singular phase transition in the present model, like in the original Ising model.

In two dimensions, the Ising model can be solved only in some specific cases \citep{onsager1944crystal} but, in general, it is known to have a critical point at some finite, nonzero temperature. On an infinite, regular, triangular lattice, the inverse critical temperature is $\beta_c = \log(3)/4$ \citep{wannier1950antiferromagnetism,zhi2009critical}. (Using a periodic square lattice does not change qualitatively the results we report here, App.~\ref{app:phase_transition_simulations}.)
In the absence of the opinion-poll term ($\varepsilon=0$), our numerical simulations on an irregular, triangular mesh ($N=1046$, Fig.~\ref{fig:phase_transition}\figlet{B}) show a similar behavior (blue dots in Fig.~\ref{fig:phase_transition}\figlet{D},\figlet{F},\figlet{H}): upon increasing $\beta$, the system acquires a spontaneous magnetization near $\beta_c$. An ordered phase then emerges, whereby the general opinion takes one of two opposed, nonzero values ($\bar{s}=\pm \eqmgn$, where $\eqmgn$ is a positive function of $\beta$).

Despite the small size of our numerical simulations ($N=1046$), the transition that occurs near the theoretical value of $\beta_c$ is sharp (Fig.~\ref{fig:phase_transition}\figlet{h}). Its rigorous characterization, however, would require larger meshes; we leave it for future investigations. Nonetheless, confident that our small simulations behave qualitatively as they should, we proceed with the introduction of opinion polls.

When the opinion-poll term  is switched on ($\varepsilon>0$), it causes a drastic change in the ordered phase ($\varepsilon=1$, pink dots in Fig.~\ref{fig:phase_transition}\figlet{D},\figlet{F},\figlet{H}). Like in one dimension, the energy density and the clustering coefficient are barely affected by $\varepsilon$ (a slight shift is visible at high $\beta$), but $\bar{s}$ remains small at any temperature (Fig.~\ref{fig:phase_transition}\figlet{H}). Above $\beta_c$, two seemingly ferromagnetic domains coexist, at the energetic cost of maintaining an interface between them---a cost that the opinion-poll term needs to balance. The next section is devoted to this energy trade-off.

\section{Continuous approximation}\label{sec:mean_field}

We now assume that the temperature is low, and that the population of voters is ordered: it has either reached a consensus, or split into two camps. We can then represent it as a connected block of $n$ voters which support one opinion, and another, opposed block of $N-n$ voters (the size of either block can vanish). These blocks are essentially uniform, but a small proportion $p$ of isolated voters can oppose the bulk of their own group ($p$ is likely to increase with temperature).

In App.~\ref{app:continuous}, we propose a continuous model which approximates the Hamiltonian, Eq.~\eqref{eq:H}; here, we briefly present the assumptions it is based on, and their implications. In the thermodynamic limit ($N\to\infty$), and in two dimensions, the two blocks are separated by a smooth interface of length $L$. This interface has an energetic cost, since it causes some voters to oppose their neighbors. On a two-dimensional triangular lattice, the number of links $n_i$ cut by this interface is, roughly, twice the number of triangles that it crosses (App.~\ref{app:interface_energy}):
\begin{equation}
  n_i = 2 \sqrt{ \dfrac{ N }{ \cal A } } L \, ,
  \label{eq:n_i}
\end{equation}
where $\cal A$ is the total area occupied by the population of voters. This area is arbitrary, since only connectivity matters, but it provides a scale for $L$.

This scaling is critical to the continuous approximation. On a regular square grid, for instance, only the prefactor of Eq.~\eqref{eq:n_i} would change, and the continuous approximation would hold, although the anisotropy of the grid would probably affect the orientation of the interface. Conversely, on a scale-free network, the continuous approximation cannot hold, and the split-society phase might altogether disappear. On which class of meshes can this phase form? This is yet an open question.

Returning to our triangular mesh, we need to estimate the average energy associated to a link cut by the interface. In general, there is no easy solution to this problem, because the two phases include mavericks voters, the proportion of which, $p$, increases near the critical temperature. When $p$ becomes significant, the very definition of the interface becomes ambiguous; we avoid this issue by assuming that the temperature is low enough for mavericks to be ignored ($p = 0$ and $\eqmgn = 1$). The energy cost associated to the boundary is then simply $2 n_i$.

The next significant assumption is that, at low temperature, the only relevant configuration is that which minimizes the total energy. We thus neglect the entropic contribution of the fluctuations, in the spirit of the mean-field theory. This turns our problem into a purely geometrical optimization: For a given value of the general opinion $\bar{s}$, we look for the configuration that minimizes $L$ (App.~\ref{app:H_continuous}). Just like intuition suggests, the best possible shape for the interface is a circle which intersects the boundary of the lattice at a right angle (or a collection of such circles). The relation between the length of the optimal interface, $L_{\mathrm{min}}$, and the general opinion, $\bar{s}$, depends on the geometry of the lattice. (Similarly, the interface between two immiscible liquids adjusts to the shape of the container.)

To fix ideas, we represent the entire population with a disk of radius unity---the geographical equivalent of a physicist's spherical cow (App.~\ref{app:opt_shape}). Elementary trigonometry then yields an implicit formula which relates the minimal length of the interface $L_{\mathrm{min}}$ to the general opinion $\bar{s}$. Accordingly, we propose a continuous approximation for the Hamiltonian, which we write as a function of $\bar{s}$ (App.~\ref{app:landscape}):
\begin{equation}
  {\cal H}_c \left(\bar{s}\right) =  \varepsilon N \left( \dfrac{4 }{ \varepsilon \sqrt{ N \pi }} L_{\mathrm{min}}(\bar{s}) + \bar{s}^2 \right) \, .
  \label{eq:Hamiltonian_mf_no_H}
\end{equation}
The state of the electorate is then found by minimizing this expression with respect to $\bar{s}$.

\begin{figure}[t]
  \centering
  \includegraphics[width=.99\linewidth]{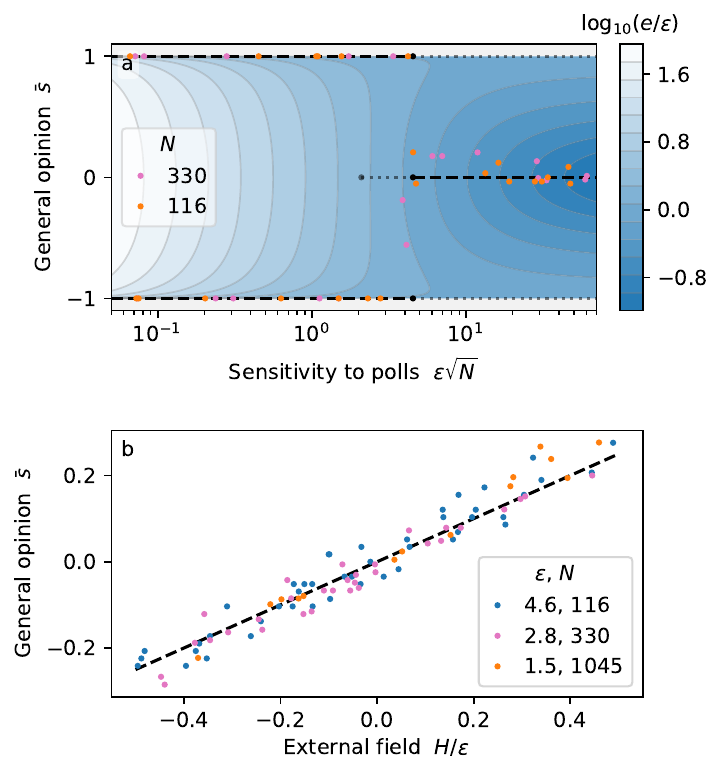}
  \caption{\figlet{A} Phase transition at low temperature. Blue shading: rescaled energy density $e/\varepsilon$, Eq.~\eqref{eq:Hamiltonian_mf_no_H}. Black dashed lines: global energy minima with respect to the average opinion $\bar{s}$. Shaded dotted lines: local minima. Dots: numerical simulations on a two-dimensional triangular mesh ($\beta = 0.5$).\label{fig:phase_transition_N}
    \figlet{B} Magnetization induced on a two-dimensional, triangular mesh by an external field $H$. The population of voters is in the split-society state. Dashed black line: continuous approximation, Eq.~\eqref{eq:susceptibility}.\label{fig:susceptibility}}
\end{figure}

Figure~\ref{fig:phase_transition_N}\figlet{A} shows the rescaled energy density $e/\varepsilon$ as a function of the general opinion $\bar{s}$, and the rescaled sensitivity to polls $\varepsilon \sqrt{ N }$. There are three energy valleys in the $(\varepsilon \sqrt{ N }, \bar{s})$ plane (loci of local minima with respect to $\bar{s}$). Two of them, along $\bar{s}=\pm1$, correspond to unanimity, whereas the the third one, along $\bar{s}=0$, represents the split-society state, which exists only when $\varepsilon \sqrt{ N } > 3 \pi^{3/2}/8$. For small values of $\varepsilon\sqrt{N}$, the two unanimity states are global minima for the energy. As $\varepsilon \sqrt{ N }$ increases, however, the system undergoes a first-order transition and, when $ \varepsilon \sqrt{ N } = 8/\sqrt{\pi}$, the split-society state becomes the global minimum (App.~\ref{app:landscape}). The parameter that controls the stability of each phase is thus $\varepsilon \sqrt{ N }$, which means that the existence of the split-society phase depends not only on the sensitivity to the polls, but also on the size of the electorate.

These results are based on an approximate, continuous model. To check them, we use a series of numerical simulations on triangular lattices (App.~\ref{app:numerical_simulations}); we find good agreement with the continuous model (Fig.~\ref{fig:phase_transition_N}\figlet{A}). The numerical simulations appear to switch from unanimity to the split-society phase when $N$ exceeds the transitional number of voters, namely
\begin{equation}
  N_t = \dfrac{64}{ \pi \varepsilon^2 } \, .
  \label{eq:N_t}
\end{equation}
Despite this good agreement, the above expression should be treated with caution. Its prefactor, in particular, depends on the geometry of the lattice. The scaling relation $N_t\sim \varepsilon^{-2}$, however, is likely to hold in other networks with a similar topology.

In summary, the split-society phase is thermodynamically stable only when voters are sensitive enough to opinion polls. Because this phase involves an interface, its stability depends on the size of the electorate; we shall make use of this dependence in Sec.~\ref{sec:pop_size}, which is devoted to actual elections. Before we do so, however, we first need to consider fluctuations.

\section{Susceptibility and fluctuations}\label{sec:susceptibility}

Thermal systems fluctuate, and the amplitude of their fluctuations depends on their sensitivity to external forces---their susceptibility. Identifying, let alone measuring, some external forcing in an electorate is, at best, challenging, but measuring fluctuations should be straightforward.

Establishing the fluctuation-dissipation relation for the Ising model has become a textbook exercise, and the opinion-poll term does not affect this classical derivation (App.~\ref{app:susceptibility}). It begins with the  introduction of an additional term, $- H N \bar{s}$, to the Hamiltonian of Eq.~\eqref{eq:H}, where $H$ represents the intensity of an external magnetic field. Here, this term serves a mathematical purpose, but one can also propose a sociological interpretation for it. It could represent a general preference for one of the two candidates, or the successful propaganda of one party. Alternatively, the polls could be biased, and systematically evaluate the general opinion to $\bar{s}-H/\varepsilon$ instead of $\bar{s}$ \citep{zhenkun2021polls}; this would add the same term to the Hamiltonian.

Clearly, the equilibrium configuration of the electorate will depend on $H$, and so will the general opinion $\bar{s}$. Assuming the system has reached the Boltzmann equilibrium, we find that the susceptibility $\chi_0$ reads (App.~\ref{app:susceptibility}):
\begin{equation}
  \chi_0 \equiv \left. \dfrac{\partial \bar{s}}{\partial H} \right|_{H=0} = \beta N \sigma_o^2 \, ,
  \label{eq:fluctuations}
\end{equation}
where $\sigma_o^2$ is the variance of the voters' opinion---the amplitude of the fluctuations. Written in this form, the susceptibility of the electorate seems unrelated to $\varepsilon$; in reality, the sensitivity to polls will express itself through $\sigma_o$.

To see this, we leave the realm of exact derivations, and return to the continuum approximation of Sec.~\ref{sec:mean_field}. Adding an external forcing, $- H N \bar{s}$, to the approximate Hamiltonian of Eq.~\eqref{eq:Hamiltonian_mf_no_H} shifts the split-society valley vertically in Fig.~\ref{fig:susceptibility}\figlet{A}. Elementary calculus yields
\begin{equation}
 \chi_0 = \dfrac{1}{2\varepsilon}
 \label{eq:susceptibility}
\end{equation}
in the thermodynamic limit ($N\to\infty$, App.~\ref{app:susceptibility}). To validate this expression, we run numerical simulations on triangular lattices; we find good agreement between the two (Fig.~\ref{fig:susceptibility}\figlet{B}). Again, the numerical prefactor in Eq.~\eqref{eq:susceptibility} depends on the topology of the lattice, but we expect the scaling $\chi_0\sim 1/\varepsilon$ to be generic.

Equation~\eqref{eq:susceptibility} shows how exotic the split-society phase is. Like a magnet below the Curie temperature, it is an ordered phase wherein the interactions between neighbors overcome thermal fluctuations. Unlike the classical ferromagnetic phase, however, its susceptibility remains finite---in the fashion of the paramagnetic phase. In that sense, the split-society phase is a hybrid.

In electoral terms, the susceptibility tells us how easily one political camp can expand at the expense of the other. For instance, if the polls are biased, the camp that appears to lose will grow, because voters will tend to oppose the winner. The actual opinion will thus shift in proportion to the susceptibility of the split-society phase. Through Eq.~\eqref{eq:fluctuations}, this susceptibility should relate the fluctuations of electoral results to the size of the electorate.

\section{Population size}\label{sec:pop_size}

Opinion polls provide some insight about the fluctuations of public opinion during an election campaign (Fig.~\ref{fig:Polish}\figlet{A}) \citep{wlezien2017dynamics}. In particular, we can measure their variance $\sigma_p$ but, because pollsters survey only a small subset of the electorate, $\sigma_p$ is not the variance of the opinion, $\sigma_o$. To distinguish the two, we average the polls over time, until their variance reaches a plateau, which we interpret as $\sigma_o$ (App.~\ref{app:op_polls}). During the 2020 Polish presidential election, for instance, we find $\sigma_o \approx 0.01$ ($0.5$ pp). This can only be an order-of-magnitude estimate, which we now use to illustrate how the present model can be compared to observations.

Combining the two expressions of the susceptibility derived in Sec.~\ref{sec:susceptibility}, Eqs.~\eqref{eq:fluctuations} and \eqref{eq:susceptibility}, we find
\begin{equation}
  \beta \varepsilon = \dfrac{1}{2N\sigma^2_o} 
  \label{eq:beta_epsilon}
\end{equation}
when the electorate is in the split-society state. In 2020, $16.4$ million voters took part in the Polish presidential election; we therefore estimate that $\beta \varepsilon \approx 3\cdot 10^{-4}$. To separate the $\varepsilon$ from $\beta$, we now use the transitional size $N_t$ of the split-society state (Sec.~\ref{sec:mean_field}).

\begin{figure}[t]
  \centering
  \includegraphics[width=.99\linewidth]{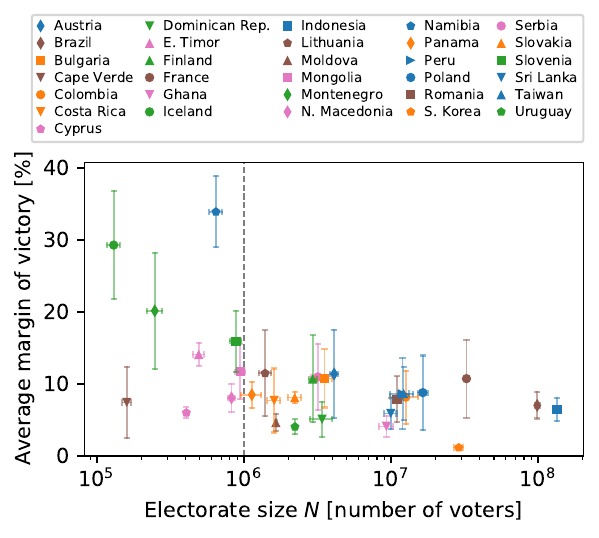}
  \caption{Average margin of victory, estimated over at least 3 binary elections since {\fieldDataDate}. Error bars show standard deviation. Dashed line: estimated transitional electorate size $N_t$ (App.~\ref{app:transition_size}).\label{fig:phase_transition_field_data}}
\end{figure}

If the social temperature $1/\beta$ is low enough, the electorate should be in an ordered state, and prone to the phase transition of Fig.~\ref{fig:phase_transition_N}\figlet{A}. Accordingly, countries with a small population would reach a near consensus during an election, whereas larger ones would find themselves in the split-society state. To check this, we collect the result of 168 binary elections in 31 countries since {\fieldDataDate}  (App.~\ref{app:election_data}). Countries where the electorate is less than about a million voters tend to generate large margins of victory (Fig.~\ref{fig:phase_transition_field_data}). Larger countries, on the other hand, typically generate a margin of victory of about 7 pp. A more careful analysis of the data confirms that a transitional population size $N_t=10^6$ indeed separates the data into two classes with distinct distributions (App.~\ref{app:transition_size}).

Looking at this empirical fact through the lens of the present model, we can estimate the sensitivity to polls with Eq.~\eqref{eq:N_t}; we then find $\varepsilon \approx 0.005$. This value means that, for an average voter, the result of opinion polls matters about 200 times less than their friends' and neighbors' opinion. This weak influence, however, accumulates over the electorate, and can ultimately maintain an entire population in the split-society state.

Combining this result with Eq.~\eqref{eq:beta_epsilon}, we find a temperature of $\beta \approx 0.07$, about a quarter of the critical value $\beta_c$ (Sec.~\ref{sec:fluct_phase_trans}). This estimate therefore suggests that $\beta$ is near the critical value, which means that the political decisions of voters are, on average, balanced---a voter is about as likely to be influenced by others as to make a personal decision.

These estimates should be treated as preliminary, as they are based on a small data set, and on several crude approximations (a specific connectivity, identical voters and constant parameters, to name a few). If confirmed, however, they would indicate that electorates self-adjust to criticality, through a mechanism that remains to be elucidated.

\section{Conclusion}

Inspired by election results and electoral maps, we have introduced the influence of opinion polls in a model  of interacting voters. The resulting system differs from the original Ising model only by a long-range interaction term, but this elementary change allows the split-society phase to emerge. This new phase is ordered, in the sense that most voters share their neighbors' opinion, but the general opinion is evenly split. Opinion polls can thus foster the emergence of two opposed camps.

The split-society phase has remarkable properties. In particular, its susceptibility remains finite and independent of temperature (unlike that of the classical ferromagnetic state). In addition, the interface that splits the electorate into two camps makes the system size-dependent, and tends to place itself along preexisting fault lines in the electorate.

These properties need to be better understood, and formally established. A specially enticing endeavor is to seek the exact partition function of a two-dimensional electorate in the thermodynamic limit. This would yield the critical temperature below which the split-society state exists, and explain why the susceptibility of this phase is constant (beyond the rudimentary reasoning of Sec.~\ref{sec:mean_field}). The kinship of the present model with the original Ising model suggests that these questions could be addressed with the tools of statistical physics, at least to some extent.

Does the split-society phase deserve its name? Its most striking properties certainly match our perception of what a split society is: two political camps of similar strength stand face to face, with little connections between them. To reverse an election, only a small number of marginal voters need to flip their allegiance, while the bulk of the two camps stick to their opinion.

These features are familiar to the observer of modern elections, but the present model needs a more rigorous evaluation. In Sec.~\ref{sec:pop_size}, we delineated a path towards its comparison with observations. Among the numerical estimates we propose, the transitional number of voters $N_t$, above which the split-society phase appears, is probably the most robust. It is, after all, directly based on election results, which are widely available and reliable data. In itself, though, it is not an exacting test of the model.

The evaluation of the two other parameters, the temperature $1/\beta$ and the sensitivity to polls $\varepsilon$, is more fragile. It relies on an accurate estimate of the opinion fluctuations, and on the assumption that these parameters are constant. Their value should be treated with caution, but we find it at least encouraging that $\beta$ is close to one. Extreme values, indeed, would be unrealistic: every voter would be either entirely disconnected from the others ($\beta\to0$), or completely determined by them ($\beta\to\infty$).

Finally, there is the connectivity network. We have limited ourselves to a two-dimensional geometry for simplicity, and for comparison with electoral maps. In reality, social networks are more complicated (possibly scale-free) and the notion of neighbors becomes hazy. The pursuit of the split-society phase in complex networks promises an exciting mathematical quest---one from which we might learn about ourselves.

\begin{acknowledgments}
We are indebted to A.J.C.~Ladd, D. Dantchev, A. Maciołek, and P. Malgaretti for seminal discussions. The manuscript benefited from the insightful suggestions of F.~Métivier, D.H.~Rothman and P.~Popovi{\'c}.
\end{acknowledgments}

%
%

\bigskip
\appendix

\input{./appendix/split_society_appendix.tex}

\bibliography{}

\end{document}

%% file: appendix/split_society_appendix.tex
\newlength{\FigWidth}
\setlength{\FigWidth}{.99\linewidth}
\newlength{\LargeFigWidth}
\setlength{\LargeFigWidth}{15cm}

\newcommand{\intlen}{\ell}
\newcommand{\ang}{\alpha}
\newcommand{\partfunc}{\mathcal{Z}}

\section{Polish presidential election of 2020}

\begin{figure}[t]
  \centering
  \includegraphics[width=\FigWidth]{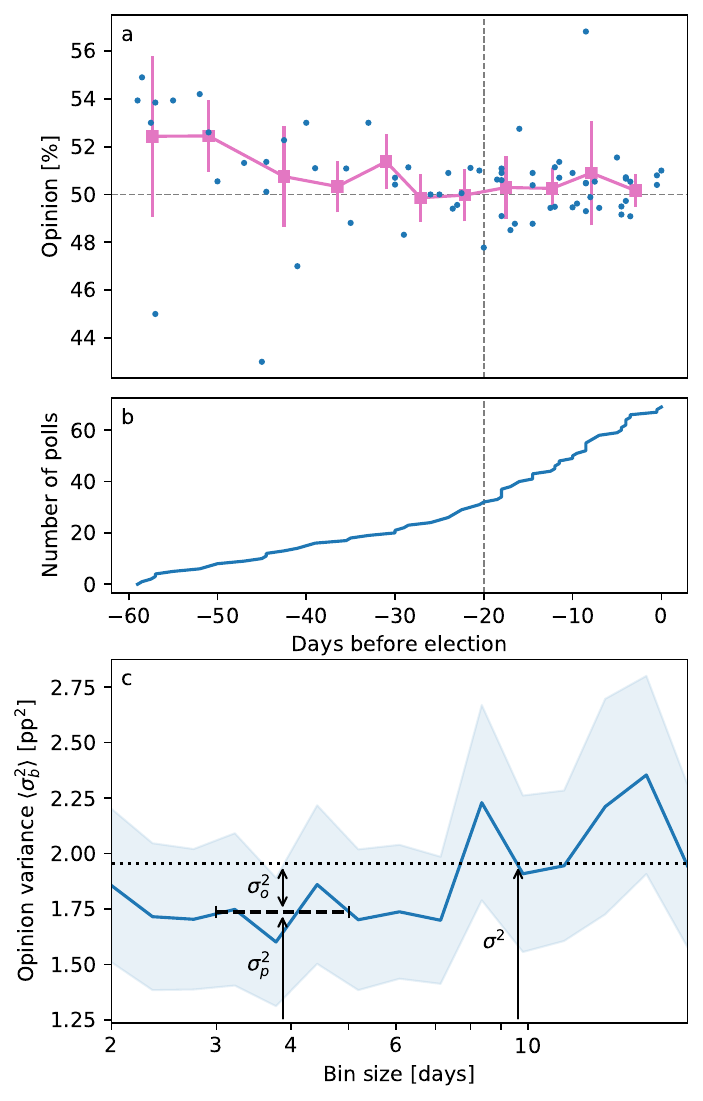}
  \caption{\figlet{A} Opinion polls before the Polish presidential election of 2020 (blue dots, in favor of Andrzej~Duda) \citep{enwiki:1059072751}. Pink squares: four-day average (error bars show the standard deviation in each bin). \figlet{B} Poll variance as a function of bin size (solid blue line, in squared percentage points). Shaded area shows variability, estimated by bootstrapping ($3/4$ of data points selected randomly 1000 times). Black dashed line: estimate of poll noise $\sigma_p^2$. Black dotted line: poll variance over entire, non-binned sample $\sigma^2$. \figlet{C} Number of opinion polls accumulated during the election (blue line). Polls less than 20 days before election appear in \figlet{B} (vertical dashed line in \figlet{A} and \figlet{C}).
  \label{fig:polls_data_SI}
  }
\end{figure}

%
%
%

\subsection{Opinion polls\label{app:op_polls}}

The 2020 presidential election in Poland lasted from June 10 to July 12. During the first round (until June 28), two candidates were selected for the vote of the second round, which took place two weeks later. Various polling organizations started probing the Polish opinion about this configuration (A.~Duda vs. R.~Trzaskowski) as early as May 14; \citet{enwiki:1059072751} have collected this data, which we use here to probe the dynamics of a binary election (Fig.~\ref{fig:polls_data_SI}\figlet{A}).

In total, 70 opinion polls make up this data set (we drop an early poll from 2019); they are distributed over the 59 days before the election. The polling rate is about 0.85 poll$\,$day$^{-1}$ initially. It increases to 2.0 poll$\,$day$^{-1}$ during the last 20 days before the second round (Fig.~\ref{fig:polls_data_SI}\figlet{b}). Since the opinion appears to have converged at the time of this transition, we limit our analysis to the polls made less than 20 days before the election.

After relaxing to a value close to 50$\,$\%, the opinion polls fluctuate around their average. We now try to separate, in this fluctuation, the contribution of measurement noise from the actual variability of the opinion.

\subsection{Fluctuations\label{app:fluctuations}}

Over the last 20 days of the election, the standard deviation of the polls is $\sigma \approx 1.40$ percentage point (pp), slightly less than the measurement uncertainty we would expect in a sample of 1000 voters, which is typical for opinion polls \cite{jackman2005pooling}. This observation alone tells us that estimating the fluctuations of a population's opinion based on surveys is fraught with statistical traps.

Indeed, assuming that opinion polls are an unbiased, but noisy, measure of the actual opinion, their variance, $\sigma_p^2$, and that of the opinion, $\sigma_o^2$, should add up to the total variance:
\begin{equation}
  \sigma^2 = \sigma_p^2 + \sigma_o^2 \, .
\end{equation}
The opinion polls of Fig.~\ref{fig:polls_data_SI}\figlet{A} provide a direct estimate of the first term only---which we find to be comparable to what we expect $\sigma_p^2$ to be for a typical opinion survey.

To distinguish the fluctuations of the polls from those of the opinion, we assume that the latter are correlated in time, whereas the former are not. This assumption is consistent with the Glauber dynamics: When the temperature is low enough (large $\beta$), only a small fraction of the population will change their mind between two polls.

If this assumption is true, and if the polls are frequent enough, we should be able to reveal the slow evolution of the opinion by assigning the data points to time bins of varying size. This binning should allow us to reduce the variability within each bin, and thus strip the time series from the polling noise.

In practice, we calculate the variance within each bin, $\sigma_b^2$, and plot its average over bins, $\langle \sigma_b^2 \rangle$, as a function of the bin size (Fig.~\ref{fig:polls_data_SI}\figlet{B}). To estimate the uncertainty about this quantity, we randomly pick three quarters of our data, and bin the data again. Repeating this bootstrapping procedure 1000 times, we find a standard deviation of about 0.4$\,$pp (20$\,$\% relative uncertainty).

As the bin size increases, the average variance in a bin also increases (the variance of a bin that contains a single data point vanishes). When the bin size reaches about three days, the average variance $\langle \sigma_b^2 \rangle$ seems to plateau around 1.75$\,$pp$^2$, before it finally reaches the total variance $\sigma^2$, after about 7 days. We interpret the first plateau as the variance of the polls $\sigma_p^2$. The rest, $\sigma^2-\sigma_p^2$, provides us with an estimate of the opinion fluctuations: $\sigma_o \approx 0.5\,$pp (expressed in terms of the variance of $\bar{s}$, $\sigma_o \approx 0.01$).

For the sake of consistency, we finally distribute the data into 5-days bins (pink line in Fig.~\ref{fig:polls_data_SI}\figlet{A}, also in Fig.~\ref{fig:Polish_map}, and calculate the variance of this binned data set---hoping that it is rid of most of the measurement noise. We then find $\sigma_o \approx 0.2\,$pp.
%
%
%


\section{Numerical simulations\label{app:numerical_simulations}}

\subsection{Numerical procedure}

To illustrate the theory and evaluate our approximations, we run  simulations on one- and two-dimensional lattices. The triangular lattices shown in Figs.~\ref{fig:evolution} and \ref{fig:phase_transition}, and used in the simulations of Fig.~\ref{fig:susceptibility}, are generated with the mesher of FreeFem++ \cite{MR3043640}. We then generate the connectivity matrix $\mathbf{J}$ associated to these meshes, by setting $J_{ij} = J_{ji}=1/2$ when nodes $i$ and $j$ are connected, and $J_{ij} = J_{ji}=0$ otherwise. The resulting matrix is sparse, since only neighboring nodes can be connected. The circular mesh of Fig.~\ref{fig:evolution}\figlet{B} has 94 nodes. The boundary of the peanut-shaped mesh of Fig.~\ref{fig:evolution}\figlet{G},\figlet{H} is parameterized by
\begin{equation}
  r = 1 + 0.7 \cos( 2 \theta ) \, ,
\end{equation}
where $r$ and $\theta$ are the polar coordinates of a point along the boundary. The mesh has 335 nodes.

We use the Glauber algorithm to evolve the state of the system \citep{glauber1963time}. At each time step, after node $i$ has been picked randomly, the entire Hamiltonian is evaluated to get the transition energy $\Delta E_i$. Clearly, this algorithm is not optimized for performance, but it serves its illustrative purpose well. All numerical routines are available online \footnote{\url{https://github.com/odevauchelle/thermocracy}}.

\subsection{Fluctuation-induced phase transition\label{app:phase_transition_simulations}}

\begin{figure}[t]
  \centering
  \includegraphics[width=\FigWidth]{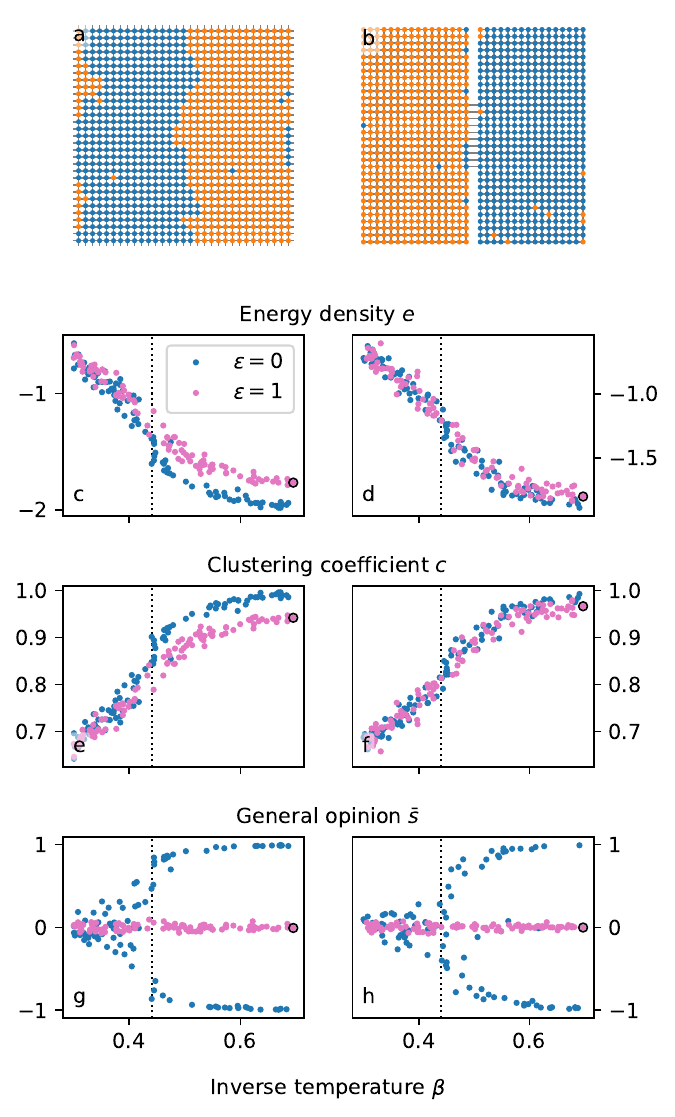}
  \caption{Fluctuation-induced transition. Left column: periodic square lattice ($N=1024$). Right column: square, non-periodic grid with bottleneck ($N=961$).
    \figlet{A}, \figlet{B} Examples of equilibrium states.
    \figlet{C}--\figlet{H} Final stage of the numerical simulation. Blue dots: no influence of the polls ($\varepsilon=0$), pink dots: polls matter ($\varepsilon=1$). Black circles correspond to states shown in \figlet{A}, \figlet{B}. The dotted black lines in  show the critical temperature for an infinite square lattice, $\beta_{c,s}=\log \left( 1 + \sqrt{2} \right)/2$ \citep{onsager1944crystal}.
  \label{fig:phase_transition_grid}}
\end{figure}

The one-dimensional lattice of Fig.~\ref{fig:phase_transition}\figlet{A} is made of 1000 nodes, each connected to exactly two neighbors. For each run, $\log_{10} \beta$ is randomly picked from a uniform distribution between $-1$ and $1$. Each simulation runs until $t=10^3$.
The two-dimensional, triangular lattice of  Fig.~\ref{fig:phase_transition}\figlet{B} has 1046 nodes; $\log_{10} \beta$ is randomly picked between $-\log_{10} 4$ and $\log_{10} 4$. Simulations run until $t=10^4$.

To check whether this transition occurs in the traditional two-dimensional Ising model, we run similar simulations on a periodic square grid (Fig.~\ref{fig:phase_transition_grid}\figlet{a}). When the influence of the polls is switched off ($\varepsilon=0$, blue dots in Fig.~\ref{fig:phase_transition_grid}\figlet{a},\figlet{c},\figlet{e},\figlet{g}), we recover the classical phase transition \citep{onsager1944crystal}. When polls matter ($\varepsilon=1$, pink dots), a similar transition seems to occur, but the general opinion $\bar{s}$ (i.e. the average magnetization) of the ordered phase vanishes. This is the signature of the split-society phase.

In the split-society phase that occurs on a periodic square grid (Fig.~\ref{fig:phase_transition_grid}\figlet{a}), the boundary between the two domains can lie anywhere on the grid. To break this symmetry, and reproduce the narrow-necked mesh of Fig.~\ref{fig:evolution}\figlet{g}, \figlet{h}, we now consider a square grid wherein a series of neighboring nodes are disconnected (Fig.~\ref{fig:phase_transition_grid}\figlet{b}). The disconnected nodes separate the grid into two equal parts, joined only through a bottleneck around the center of the grid. This topology barely affects the phase transition (Fig.~\ref{fig:phase_transition_grid}\figlet{d},\figlet{f},\figlet{h}) but, this time, the two domains of the split-society phase tend to occupy each side of the bottleneck, thus minimizing the length of the boundary between them.

\subsection{First-order transition at low temperature}

The simulations shown on Fig.~\ref{fig:susceptibility} were made on three triangular meshes of circular shape, with 116, 330 and 1045 nodes. In Fig.~\ref{fig:susceptibility}\figlet{A} the temperature is fixed to $\beta = 0.5$, while $\log_{10} (\varepsilon\sqrt{N} )$ is randomly picked between $\log_{10}7 - 2$ and $\log_{10}7 + 1 $. Each simulation runs until $t=10^3$. In Fig.~\ref{fig:susceptibility}\figlet{B}, $\beta = 1$ and $\varepsilon$ is fixed to the value indicated in the legend. The (rescaled) exterior magnetic field $H/\varepsilon$ is picked randomly between $-1/2$ and $1/2$. Each simulation is run until $t=10^3$.


\section{Continuous limit\label{app:continuous}}

\subsection{Political camps\label{sec:thermodynamic_limit}}

We consider the Hamiltonian of Eq.~\ref{eq:H} in the limit of a large number of voters ($N \to \infty$), when they are distributed on a two-dimensional lattice, such as the one of Fig.~\ref{fig:phase_transition}\figlet{B}. In the presence of an external field $H$, this Hamiltonian becomes
\begin{equation}
  {\cal H} = - \sum_{i,j} J_{ij} s_i s_j + \dfrac{\varepsilon}{N} \left( \sum_i s_i \right)^2 - H \sum_i s_i \, .
  \label{eq:Hamiltonian}
\end{equation}

We want to approximate our system as a two-dimensional continuum. {To do so, we first consider that the voters distribute themselves over a fixed domain. The density of voters thus grows with $N$, and we can introduce the local opinion $s$, which is some local average of the opinions of nearby voters. By construction, in the thermodynamic limit ($N \to \infty$), $s$ becomes a function of position which can take any real value between $-1$ and $+1$.}

When the temperature is low enough, we expect that the population of voters will form one or two homogeneous phases. In other words, the disk that {represents} the population will be either covered with a single phase of {local opinion $+\eqmgn$ or $-\eqmgn$} (Fig.~\ref{fig:evolution}\figlet{D}), or split into two phases of opposite {opinions} {$+\eqmgn$ and $-\eqmgn$} (Fig.~\ref{fig:evolution}\figlet{F}). {This allows us to introduce the next simplification: We assume that the local opinion $s$ can take only two values, $+\eqmgn$ and $-\eqmgn$. The local opinion $s$ thus splits the main domain into subdomains with distinct phases (political camps).}

Each phase represents a group of connected voters, most of whom share a common opinion. Within each phase, however, there are isolated voters whose opinion oppose the majority of their group. As a result, ${\eqmgn}$ can be less than 1, especially when the temperature gets {close to its critical value}. In fact, we expect these phases to be essentially the same as those of the classical Ising model ($\varepsilon=0$) at low temperature.

We now return to the Hamiltonian of the original model, Eq.~\eqref{eq:Hamiltonian}. Its first term accounts for the interactions between neighbors.  In the continuous limit, and for an isotropic and homogeneous mesh, we expect this contribution to be proportional to the length $
\intlen$ of the interface separating two phases; we shall estimate this term in the next  section. As for the energy stemming from the interaction of neighboring voters within each phase, it is independent from the shape of the political camps, and its density is the same in both camps. It can thus be discarded as a constant contribution to the total energy.

The second and third terms of the Hamiltonian depend on the system's state only via the global opinion 
\begin{equation}
\bar{s}
  = \dfrac{1}{N}\,\sum_{i=1}^N s_i 
  = \eqmgn \left( \dfrac{2A}{\cal A} - 1 \right)\, ,
  \label{eq:s_of_A}
\end{equation}
where $A$ is the area occupied by the phase of local opinion $\eqmgn$, and ${\cal A}$ is that of the whole domain. As a consequence, the system's energy depends essentially on the area of one of the phases ($A$, for instance) and on the length of the interface $\intlen$. Therefore, we can use these two parameters (or, equivalently, $\bar{s}$ instead of $A$) to describe the system---this is our third, and last, simplification. Of course, these parameters are not fully independent; when $A$ vanishes, for instance, $\intlen$ must also vanish.

\subsection{Interface energy\label{app:interface_energy}}

\begin{figure}[t]
  \centering
  \includegraphics[width=\FigWidth]{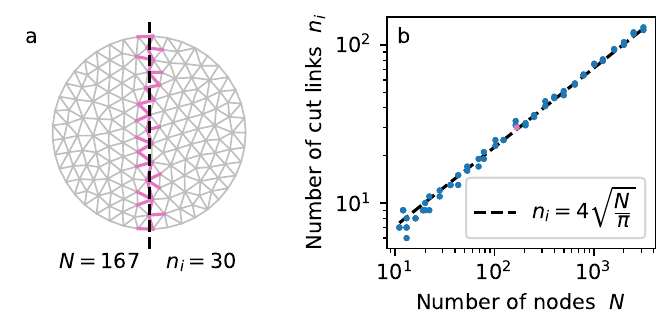}
  \caption{Interface between two phases. \figlet{A}:~Example of straight interface (dashed black line) cutting through a triangular mesh (grey lines). The number of edges cutting the interface is $n_i$ (pink lines). \figlet{B}:~Relation between the number of nodes $N$ and the number of edges crossing the interface, $n_i$ (blue dots). The pink dot corresponds to panel~\figlet{A}. The black dashed line shows Eq.~\eqref{eq:n_i}.
  \label{fig:cut_links}
  }
\end{figure}

To calculate the energy increase associated to the interface between two phases, we first need to estimate the number of links that are crossed by this interface, and then the energy associated to each link.

If, on average, the mesh is homogeneous and isotropic, the number of links that an interface crosses, $n_i$, does not {depend} on the location of the interface, nor on its orientation. In the thermodynamic limit ($N\to \infty$), the typical distance $d_N$ between nodes becomes infinitesimal. Therefore, if the interface is smooth, it will appear as a straight line at the scale of $d_N$. In other words, $n_i$ depends only on the interface's length $L$, not on its shape. We can thus calculate $n_i$ along any smooth path we like.

Let us consider a path that runs straight through an isotropic and homogeneous triangular mesh, such as the one of Fig.~\ref{fig:cut_links}\figlet{A}. On average, this path will cut two edges over a distance $d_N$ (the average distance between nodes). This suggests $n_i \approx 2 L/d_N$. To estimate $d_N$, we simply take the square root of the inverse density of nodes: $d_N\approx \sqrt{ {\cal A}/N } $. Overall, we find Eq.~\eqref{eq:n_i}.
To test this estimate, we generate a series of triangular mesh over a disk of radius unity (${\cal A} =\pi$, Fig.~\ref{fig:cut_links}\figlet{A}), and count the number of edges that intersect a diameter ($L=2$). We find that Eq.~\eqref{eq:n_i} is in excellent agreement with the exact results (Fig.~\ref{fig:cut_links}\figlet{B}).

We now need to evaluate the energy gain associated to each link crossed by the interface. To do so, we first consider that the whole domain is invaded by one of the two phases. At a low enough temperature, the phase is virtually homogeneous, and the general opinion is ${\eqmgn}\approx 1$. We then draw an interface, and flip all the nodes that are on one side of the interface. For each link cut by the interface, the average energy increase is
\begin{equation}
  \delta E = 2 \, ,
  \label{eq:delta_E}
\end{equation}
whereas the energy associated to the other links does not change. When the temperature increases, we expect that there will be more and more isolated voters who oppose the opinion of their phase; ${\eqmgn}$ would then {decrease}, and so would the energetic cost of the interface. Hereafter, for simplicity, we assume that the temperature is low enough that $\eqmgn=1$.

Understanding the structure of an interface in the two-dimensional Ising model is an end in itself, which requires a rigorous, dedicated investigation \cite{Abraham1976}. The simplistic approach we use here will eventually break upon approaching the critical temperature. Before this happens, however, we expect the continuous model to behave like the original, discrete model, at least when $N$ is large and the temperature is well below criticality.

\subsection{Hamiltonian of the continuous model\label{app:H_continuous}\label{app:opt_shape}}

We can now combine equations~\eqref{eq:Hamiltonian}, \eqref{eq:n_i}, and \eqref{eq:delta_E} to write {a} Hamiltonian ${\cal H}_c$ for the continuous model:
\begin{equation}
  {\cal H}_c = 4 \sqrt{ \dfrac{N}{\cal A} } \intlen + \varepsilon N \bar{s}^2 - H N \bar{s} \, .
  \label{eq:Hamiltonian_mf}
\end{equation}
where one can use Eq.~\eqref{eq:s_of_A} to replace $\bar{s}$ with $A$.

From now on, we shall interpret this expression within the framework of the mean-field theory. In other words, we shall look for a configuration that minimizes the Hamiltonian ${\cal H}_c$ (as opposed to the free energy of the system). To do so, we need to relate the length $\intlen$ of the interface to the general opinion $\bar{s}$ or, equivalently, to the area $A$ of one of the two political camps.

\begin{figure}[t]
  \centering
  \includegraphics[width=\FigWidth]{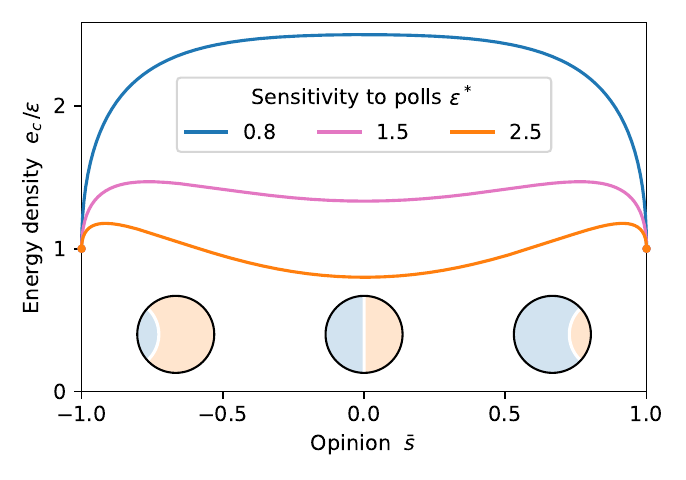}
  \caption{Energy landscape for a population of voters at low temperature. The three curves represent Eq.~\eqref{eq:E_mf} with $H=0$. The colored disks show the geometrical distribution of the opposed phases ($\bar{s}$ corresponds to the center of each disk).
  \label{fig:mean_field_energy}}
\end{figure}

At low temperature, the population of voters will tend to minimize its total energy, which we can treat as an approximation of its free energy. If the approximations of section~\ref{sec:thermodynamic_limit} hold, we need to minimize the Hamiltonian ${\cal H}_c$ with respect to the general opinion $\bar{s}$ and the length ${\intlen}$ of the interface. Fortunately, we can minimize the Hamiltonian in two steps: we first look for the minimum {value of $\intlen$ for a fixed value of $\bar{s}$ (and thus $A$), and then minimize the resulting Hamiltonian with respect to $\bar{s}$.

The first step of this procedure is a purely geometrical problem, which is analogous to finding the shape of a (two-dimensional) droplet of oil in water: surface tension minimizes the length of the interface between the two fluids. The droplet's shape depends on the container's, and on the contact angle of the triple point (where the interface joins the container). Here, for simplicity, we assume that the domain is a disk of unit radius, and therefore of area ${\cal A}=\pi$ (Fig.~\ref{fig:L_A_relation}\figlet{A}). Our aim is then to find the shape that minimizes the length ${\intlen}$ of the interface between two {phases} on this disk, for a fixed value of $A$ (the area of the first phase).

Based on the analogy with surface tension, we know that this optimal droplet will be a portion of a disk (Fig.~\ref{fig:L_A_relation}\figlet{A}). This observation reduces our optimization problem to two parameters: the radius of the droplet, and the distance from its center to the {center of the system}.

Our Hamiltonian is insensitive to the interface between any of the two phases and the outside world (the border of the country, black line in Fig.~\ref{fig:L_A_relation}\figlet{A}). Pursuing the surface tension analogy further, this observation corresponds to neutral wettability, which implies that the contact angle at the triple point is $90^\circ$. Our optimization problem now depends only on a single parameter, which we chose to be the angle ${\ang}$ formed by the centers of the two disks and one of their intersections (Fig.~\ref{fig:L_A_relation}).

We now need to identify the circles that {intersect}  the unit disk at a normal angle, and determine the area $A$ of the intersection and the length $ {\intlen} _{\mathrm{min}}$ of the interface. Based on elementary geometry, we find that the two quantities can be expressed as functions of $\ang$:
\begin{equation}
  A = \dfrac{\pi}{2} -  {\ang}  + \dfrac{ {\ang} }{( \tan \ang )^2} - \dfrac{1}{\tan \ang}
  \label{eq:A}
\end{equation}
\begin{equation}
  \intlen_{\mathrm{min}} = \dfrac{2 \ang}{\tan \ang} \, .
  \label{eq:L}
\end{equation}
We have thus parameterized the relation $\intlen_{\mathrm{min}}(A)$ (Fig.~\ref{fig:L_A_relation}\figlet{B}). As expected, in the split-society state ($\bar{s}=0$, $A=\pi/2$, $\ang \to 0$), the length of the interface is $\intlen_{\mathrm{min}}=2$, that is, the diameter of the unit disk.

\begin{figure}[t]
  \centering
  \includegraphics[width=\FigWidth]{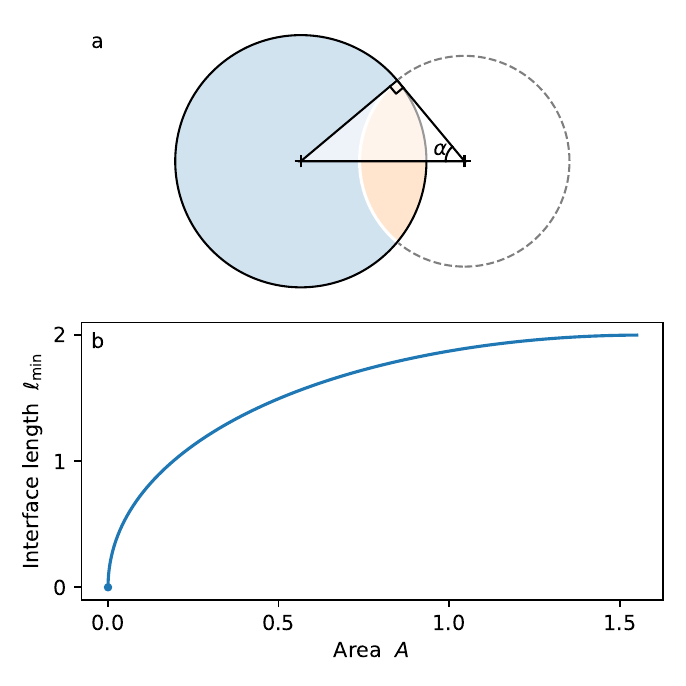}
  \caption{Optimal droplet in a disk. \figlet{A}:~Example of optimal droplet (orange area), for $\bar{s}=3/4$ (equivalently, $A=\pi/8$). The small black square indicates a right angle. \figlet{B}:~Relation between the optimal length $\intlen_{\mathrm{min}}$ and the droplet area $A$, equations~\eqref{eq:L} and \eqref{eq:A}.
  \label{fig:L_A_relation}
  }
\end{figure}

\subsection{Energy landscape\label{app:landscape}}

We are now ready to express the energy of the population as a function of the general opinion $\bar{s}$ only. To do so, we note that for a configuration to minimize the Hamiltonian~\eqref{eq:Hamiltonian_mf}, it needs to minimize the length $\intlen$ of the interface between the two phases.  Thus, using relation~\eqref{eq:s_of_A}, the energy density $e_c$ of the population reads, in the mean field theory,
\begin{equation}
  \dfrac{e_c}{ \varepsilon } =
    \dfrac{1}{ \varepsilon^* } \, \intlen_{\mathrm{min}} \left( \dfrac{\pi}{2} \left(  1 + \bar{s} \right) \right)
    + \bar{s}^2
    - \dfrac{H}{\varepsilon} \bar{s} \, ,
  \label{eq:E_mf}
\end{equation}
where we have defined the (rescaled) sensitivity to polls as
\begin{equation}
\varepsilon^* = \dfrac{ \varepsilon \sqrt{ N \pi }}{4}   \, .
\end{equation}
To convert Eq.~\eqref{eq:E_mf} into Eq.~\ref{eq:Hamiltonian_mf_no_H}, we simply need to choose $H=0$ and, for convenience, define
\begin{equation}
L_{\mathrm{min}}(\bar{s})=\intlen_{\mathrm{min}} \left( \dfrac{\pi}{2}\, \left(  1 + \bar{s} \right) \right) \, .
\end{equation}

Let us first consider Eq.~\eqref{eq:E_mf} in the absence of any external field ($H=0$, Fig.~\ref{fig:mean_field_energy}). When the sensitivity of the population to polls is small enough (blue curve, Fig.~\ref{fig:mean_field_energy}), the energy reaches a maximum for $\bar{s}=0$ (split-society state), and two minima of equal depth for $\bar{s}=\pm1$ (consensus). Above some transitional value, the maximum turns into a local minima (pink curve, Fig.~\ref{fig:mean_field_energy}).
This transition occurs when the second derivative of the energy vanishes, that is, when
\begin{equation}
  \varepsilon^* = - \dfrac{\pi^2}{8} \,  \intlen^{\prime\prime}_{\mathrm{min}}\left( \dfrac{\pi}{2} \right) \, .
\end{equation}
Using equations~\eqref{eq:A} and \eqref{eq:L}, we find that there is a local minimum at 
$\bar{s}=0$ when $ \varepsilon^* > 3\pi^2/32$. We now need to know whether the minimum for $\bar{s}=0$ is a global one. This is easily achieved by evaluating Eq.~\eqref{eq:E_mf} at $\bar{s}=0$ (then $\intlen_{\mathrm{min}}=2$), and at $\bar{s}=\pm1$ (then $\intlen_{\mathrm{min}}=0$). We find that the split-society state is a global minimum when $\varepsilon^* > 2$. This value, of course, relies on our assumption that ${\eqmgn}\approx 1$. At non-zero temperatures, the energy of the interface should decrease, and $\bar{s}$ gets bounded by $-\eqmgn$ and ${+\eqmgn}$; these changes will affect the stability of the split-society state.

The above reasoning holds only for the global} minimum of the energy (thermodynamically stable state) but, as suggested by Fig.~\ref{fig:mean_field_energy}, there could also be local minima (metastable states). A thorough investigation of the latter would require us to consider distinct droplets of the same phase, or droplets away from any boundary. As this would significantly complicate the analysis, we shall not pursue this investigation here.

\subsection{Susceptibility of the split-society state\label{app:susceptibility}}

We now turn our attention to the susceptibility of a population: How does the general opinion $\bar{s}$ change when some external forcing $H$ is applied to the electorate? We limit ourselves to the split-society case ($\bar{s}  \approx  0$). Up to second order, the population's energy reads
\begin{equation}
  \dfrac{e_c}{ \varepsilon N } \approx
    \dfrac{2}{ \varepsilon^* } + \left( 1 - \dfrac{3\pi^2}{32 \varepsilon^* }  \right) \bar{s}^2
    - \dfrac{H}{\varepsilon} \bar{s} \, .
  \label{eq:E_mf_ss}
\end{equation}
The application of an external field $H$ thus shifts the minimum of the population's energy. This minimum can be found by differentiating Eq.~\eqref{eq:E_mf_ss} with respect to $\bar{s}$, which yields the susceptibility $\chi_0$ of the split-society phase:
\begin{equation}
  \chi_0 = \left. \dfrac{\partial \bar{s} }{\partial H} \right|_{\bar{s}=0} = 
    \dfrac{1}{2 \varepsilon \left( 1 - \dfrac{3\pi^2}{32 \varepsilon^*}\right)} \,.
    \label{eq:susceptibility_th}
\end{equation}
As expected, the susceptibility diverges when the split-society phase disappears ($\varepsilon^*\to3\pi^2/32$). In the thermodynamic limit ($N\to\infty$), or when the population is very sensitive to polls (${\varepsilon}\to\infty$), the susceptibility takes the simple form
\begin{equation}
  \lim_{ \varepsilon^* \to\infty} \chi_0 = \dfrac{1}{2\varepsilon} \, .
\end{equation}
The numerical simulations shown in Fig.~\ref{fig:susceptibility}\figlet{B} are compatible with this approximate expression. In the next section, in line with classical statistical physics, we relate the susceptibility to the intensity of the fluctuations.


\section{Fluctuation-dissipation relation}

\newcommand{\eqFluct}{10}

When its evolution is governed by the Glauber dynamics, {our discrete model} relaxes to the Boltzmann equilibrium. A general consequence of this equilibrium is the relation between the amplitude of the fluctuations, and the susceptibility of the system to external forces---an instance of the fluctuation-dissipation theorem \citep[e.g.,][]{ma1985statistical}.

The Ising model we use yields this classic relation, in the form of Eq.~\ref{eq:fluctuations}. The new term that we added to the Hamiltonian, which accounts for the influence of opinion polls, does not affect the textbook derivation of the fluctuation-dissipation relation. To show this, we begin with recalling the probability density in the Boltzmann equilibrium:
\begin{equation}
  \rho(\mathbf{s}) = \dfrac{\exp \left( -\beta {\cal H}(\mathbf{s}) \right) }{\partfunc} \, ,
\end{equation}
where the partition function $\partfunc$ reads
\begin{equation}
  \partfunc = \sum_{\{\mathbf{s}\}} \exp \left({ -\beta {\cal H}(\mathbf{s}) }\right) \, .
  \label{eq:partition}
\end{equation}
The above summation applies to all possible configurations of the electorate (the values of all $s_i$). From there, it is just a matter of arithmetic to reach Eq.~\ref{eq:fluctuations}. Indeed, the opinion of the electorate reads
\begin{equation}
  \bar{s} = \dfrac{1}{N} \sum_{i} s_i \, .
\end{equation}
For later convenience, we now introduce
\begin{equation}
M=N\bar{s} \, ,
  \label{eq:mag}
\end{equation}
and call this quantity ``magnetic moment'', as is customary in the context of the original Ising model. The Hamiltonian $\cal{H}$, as expressed in Eq.~\eqref{eq:Hamiltonian}, depends linearly on the external field $H$, and the coefficient of this relation is just the magnetic moment:
\begin{equation}
  M = - \dfrac{\partial \cal{H}}{\partial H} \, .
\end{equation}
This result is entirely independent from the term we have introduced in the Hamiltonian (the impact of opinion polls). Combining it with Eq.~\eqref{eq:partition} yields
\begin{equation}
  \dfrac{1}{\partfunc} \dfrac{\partial \partfunc}{\partial H} = \beta \langle M \rangle
  \label{eq:dZ_dH}
\end{equation}
where the brackets denote the average over all configurations:
\begin{equation}
  \langle M \rangle = \sum_{\{\mathbf{s}\}} \rho ( \mathbf{s} ) \, M ( \mathbf{s} ) \, .
\end{equation}
Differentiating the partition function once more, we get
\begin{equation}
  \dfrac{1}{\partfunc} \dfrac{\partial^2 \partfunc}{\partial H^2} = \beta^2 \langle M^2 
  \rangle
  \, .
  \label{eq:dZ2_dH2}
\end{equation}
Finally, differentiating Eq.~\eqref{eq:dZ_dH} with respect to $H$, and combining the result with Eq.~\eqref{eq:dZ2_dH2} yields
\begin{equation}
  \dfrac{\partial \langle M \rangle}{\partial H} = \beta \left( \langle M^2 \rangle - \langle M \rangle^2 \right) \, ,
\end{equation}
which we can rephrase in terms of the opinion's variance and susceptibility using equations~\eqref{eq:mag} and \eqref{eq:susceptibility_th}:
\begin{equation}
  \chi = \beta N \sigma_o^2 \, .
\end{equation}
In the absence of any external field ($H=0$), $\chi$ becomes $\chi_0$, and the above formula is Eq.~\ref{eq:fluctuations}.

\section{Margin of victory}

\begin{figure}
  \centering
  \includegraphics[width=\FigWidth]{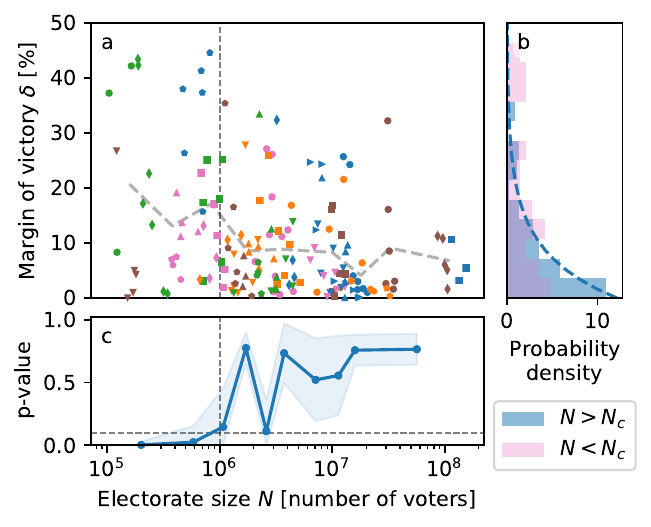}
  \caption{\figlet{A} Margin of victory in quasi-binary elections as a function of the electorate size (colored markers correspond to Fig.~\ref{fig:phase_transition_field_data}). Dashed gray line: logarithmic binning of the same data. Vertical dashed line: estimated transitional number of voters $N_t=10^6$. \figlet{B} Probability distribution of the margin of victory. Blue (resp. pink) bars: population larger (resp. smaller) than $N_t$. Blue dashed line: exponential distribution fitted to electorates larger than $5\cdot10^6$ voters. \figlet{C} Probability that the observed margin of victory for electorates smaller than $N$ is sampled from the exponential distribution of \figlet{B} (p-value from Pearson's chi-squared test over 3 bins). The shaded area shows the standard deviation estimated by bootstrapping (1000 random samples of $3/4$ of the data set). The dashed black line is $p=0.1$ ($90\,$\% confidence interval).
  \label{fig:phase_transition_SI}
  }
\end{figure}

\subsection{Election data\label{app:election_data}}

\newcommand{\EarliestDate}{1990}

To compare the present model to observations, we focus on country-wide elections. We first select countries that are either  ``full'' or ``flawed'' democracies according to the \emph{Economist} Intelligence Unit \cite{unit2021democracy}. Among those, we select countries where the head of either state or government is chosen by direct election \cite{enwiki:1099784270}. We then collect election results with a Python routine \footnote{\url{https://github.com/odevauchelle/WikipediaElection}} that parses the dedicated Wikipedia web pages \citep[e.g.][]{enwiki:1067263474}. After the result of an election is collected, the routine fetches the page of the previous election in the same country, and so on until the election occurred before {\EarliestDate} (or until the series of elections ends).

We then need to ensure that each election was a binary choice. To do so, we keep only elections for which the scores of the two leading candidates sum up to the total number of votes. (This condition is always satisfied in some countries, like Poland or France, where the second round of the presidential election involves only the two candidates who lead the first round.) Finally, we keep in our data set only countries for which there are more than 3 binary election results (this allows us to estimate, albeit roughly, the standard deviation of the result). In total, we are left with 168 elections in 31 countries.

We call $N_1$ and $N_2$ the number of votes received by the two leading candidates, and define the size of the electorate as $N = N_1 + N_2$ (the actual electorate can be a bit larger than $N$). For each election, the margin of victory is then $\delta = \vert N_1 - N_2 \vert/N$. Figure~\ref{fig:phase_transition_SI}\figlet{A} shows the margin of victory, $\delta$, as a function of the electorate size, $N$, for the complete data set.

\subsection{Probability distribution of the margin of victory\label{app:margin_distrib}}

When the electorate is larger than a few million voters, most elections are tight (Fig.~\ref{fig:phase_transition_SI}\figlet{A}). Conversely, in smaller countries (Iceland, Cape Verde and Monte Negro in our data set), many elections lead to a landslide victory or a near consensus. We now investigate this transition.

We first select the 65 elections whose electorate is larger than $5\cdot10^6$ voters (large countries), and plot the probability distribution of their result (blue bars in Fig.~\ref{fig:phase_transition_SI}\figlet{B}). The resulting histogram is well approximated by an exponential distribution $f$:
\begin{equation}
  f(\delta) = \dfrac{1}{\delta_l} \exp \left( { -\delta/\delta_l } \right)
  \label{eq:exp_distrib}
\end{equation}
where $\delta_l \approx 7.1\,$\% is the average margin of victory in large countries (dashed blue line in Fig.~\ref{fig:phase_transition_SI}\figlet{B}).

Equipped with this distribution of reference, we can now look for the subset of our data to which it does not apply. For illustration, the pink bars in Fig.~\ref{fig:phase_transition_SI}\figlet{B} show the distribution of the margin of victory in countries where the electorate is smaller than $3\cdot10^5$ voters. It features a peak around a margin of 45$\,$\%, and therefore does not look like the exponential distribution of Eq.~\eqref{eq:exp_distrib}.

\subsection{Transitional population size\label{app:transition_size}}

To locate the transition from consensus to tight elections, we look for the transitional number of voters, $N_t$, below which the results are unlikely to be drawn from the exponential distribution of Eq.~\eqref{eq:exp_distrib}. Looking at figure Fig.~\ref{fig:phase_transition_SI}\figlet{A}, we expect that this number will lie somewhere between $5\cdot10^5$ and $5\cdot10^6$ voters.

To refine this estimate, we pick a value for $N_t$, and select the election whose electorate is smaller than this value. We then distribute the results into 3 bins of equal size between 0 and 50$\,$\%. Finally, we estimate the likelihood of this 3-bins histogram based on the chi-square distribution, assuming that the probability density is Eq.~\eqref{eq:exp_distrib} (Pearson's chi-squared test). We then evaluate the corresponding p-value---the probability that the difference between the histogram and the exponential distribution is at least what we find. This p-value is represented as a function of the transitional electorate size in Fig.~\ref{fig:phase_transition_SI}\figlet{C}.

When the guessed value of $N_t$ is less than about $9\cdot10^5$ voters, the p-value is consistently less than 0.1 ($90\,$\% confidence interval). In other words, this data is unlikely to be a sample from Eq.~\eqref{eq:exp_distrib}. Conversely, this conclusion does not hold any more for larger values of $N_t$, except at about $2.5\cdot10^6$ voters. Finally, we evaluate the robustness of this result by bootstrapping (1000 random samples of $3/4$ of the data set), and find that $N_t$ lies between $6\cdot 10^5$ and $3\cdot 10^6$ voters (shaded area in Fig.~\ref{fig:phase_transition_SI}\figlet{C}, $80\,$\% confidence interval). In views of these results, we estimate $N_t$ to about a million voters.

%
%
%
%